\documentclass[showpacs]{revtex4}

\usepackage{graphicx}
\usepackage{dcolumn}
\usepackage{amsmath}
\makeatletter
\def\btt#1{\texttt{\@backslashchar#1}}
\DeclareRobustCommand\bblash{\btt{\@backslashchar}} \makeatother

\begin{document}

\title[]{Radiating black hole solutions in arbitrary
dimensions}
\author{S. G.~Ghosh}\thanks{E-mail: sgghosh@iucaa.ernet.in}
\affiliation{{ BITS, Pilani DUBAI Campus, P.B. 500022, Knowledge
Village, DUBAI, UAE and \\ Inter-University Center for Astronomy
and Astrophysics,
 Post Bag 4\\ Ganeshkhind,  Pune - 411 007, INDIA }%
 }

\author{A. K.~Dawood}
\affiliation{%
Inter-University Center for Astronomy
and Astrophysics,
 Post Bag 4\\ Ganeshkhind,  Pune - 411 007, INDIA}%

\date{\today}
\begin{abstract}
We prove a theorem that characterizes a large family of non-static
solutions to Einstein equations in $N$-dimensional space-time,
representing, in general, spherically symmetric Type II fluid. It
is shown that the best known Vaidya-based (radiating) black hole
solutions to Einstein equations, in both four dimensions (4D) and
higher dimensions (HD), are particular cases from this family. The
spherically symmetric static black hole solutions for Type I fluid
can also be retrieved. A brief discussion on the energy
conditions, singularities and horizons is provided.
\end{abstract}


\maketitle

\section{Introduction}
In recent time, it was demonstrated that the string theory requires higher dimensions
for its consistency. Models with the space-time with large extra
dimensions were recently proposed in order to solve the hierarchy
problem, that is to explain why the gravitational coupling
constant is much smaller than the coupling constants of other
physical interactions.  These new concepts of higher dimensional
physics have a number of interesting applications in modern
cosmology and theory of gravity. This has triggered,  during the
past decade, a significant increase in interest in black holes in
higher dimensions (HD) (see, for example, the review articles of
Horowitz \cite{gth} and Peet \cite{ap}). There is now an extensive
literature of solutions in string theory with horizons, which
represent black holes, and related objects in arbitrary
dimensions. The physical properties of these solutions have been
widely studied. Interest in black holes in HD has been further
intensified in recent years, due, for example, to the role they
have played in the conjectured correspondence between string
theory (or supergravity) on asymptotically locally anti-de Sitter
backgrounds and the large-N limit of certain conformal field
theories defined on the boundary-at-infinity of these backgrounds
\cite{agm,mj,ew}.

Static and spherically symmetric space-times are one of the
simplest kinds of space-times that one can imagine in general
relativity. Yet, even in this simple situation, solving the
Einstein field equations may be far from trivial. In fact, it
turns out that such a problem is intractable due to complexity of
the Einstein field equations. Hence, there are very few
inhomogeneous and nonstatic solutions known, one of them is the
Vaidya solution. The Vaidya solution \cite{pc} is a solution of
Einstein's equations with spherical symmetry for a null fluid
(radiation) source (a Type II fluid) described by energy momentum
tensor $T_{ab} = \psi l_a l_b$, $ l_a$ being a null vector field.
The Vaidya's radiating star metric is today commonly used for two
purposes: (i) As a testing ground for various formulations of the
Cosmic Censorship Conjecture (CCC). (Actually CCC is a
famous  conjecture, first formulated by
Penrose \cite{rp}. The conjecture, in it's weak version,  essentially state
that any naked singularity which is created by evolution of regular
initial data will be shielded from the external view by an event
horizon. According to the strong version of the CCC, naked
singularities are never produced, which in the precise mathematical
terms demands that space-time should be globally hyperbolic.) (ii) As an exterior solution for models of objects consisting
of heat-conducting matter. Recently, it has also proved to be
useful in the study of Hawking radiation,
 the process of black-hole evaporation \cite{rp1}, and in the
stochastic gravity program \cite{hv}.  It has also advantage of
allowing a study of the dynamical evolution of horizon associated
with a radiating black hole.

Also, several solutions in which the source is a mixture of a
perfect fluid and null radiation have been obtained in later years
\cite{ka}. This includes the Bonnor-Vaidya solution \cite{bv} for
the charge case, the Husain solution \cite{vh} with an equation of
state $P = k \rho $.  Glass and Krisch  \cite{gk} further
generalized the Vaidya solution to include a string fluid, while
charged strange quark fluid (SQM) together with the Vaidya null
radiation has been obtained by Harko and Cheng \cite{hc} (see also
\cite{ns1}). Wang and Wu \cite{ww} further extrapolated the Vaidya
solution to more general case, which include a large family of
known solutions.

Motivated by this and by a recent work Salgado \cite{ms}, we
\cite{dg} have proved a theorem characterizing a three parameter
family of solutions, representing, in general, spherically
symmetric Type II fluid that includes most of the known solutions
to Einstein field equations. This is done by imposing certain
conditions on the energy momentum tensor (EMT) (see also
\cite{vk,rg,id,eg}).

In this paper, we consider an extension of our work \cite{dg}, so that a
large family of exact spherically symmetric Type II fluid
solutions, in arbitrary dimensions, are possible, including it's
generalization to asymptotically de Sitter/anti-de Sitter.

\section{The Radiating black-hole solutions}

\noindent {\it{\bf Theorem - I}:  Let ($M,~g_{ab}$) be an N-dimensional space-time [sign$(g_{ab})$ = $+(N-2)$]
such that (i) It is non-static and spherically symmetric, (ii) it satisfies Einstein field equations, (iii) in
the Eddington-Bondi coordinates where \\ $ ds^2 = - A(v,r)^2 f(v,r)\;  dv^2
 +  2 \epsilon A(v,r)\; dv\; dr + r^2 (d \Omega_{N-2})^2$,
where, $ (d \Omega_{N-2})^2 = d \theta^2_{1} + sin^2({\theta}_1) d \theta^2_{2} + sin^2({\theta}_1)
sin^2({\theta}_2)d \theta^2_{3} + \ldots + \left[\left( \prod_{j=1}^{N-2} sin^2({\theta}_j) \right) d
\theta^2_{N-1} \right]$, the energy-momentum tensor $T^{ab}$ satisfies the conditions $T^v_r=0$ and $
T^{\theta_1}_{\theta_1} = kT^r_r$, $(k = \mbox{const.}\; \in \; \mathcal{R}$) (iv) it possesses a regular
Killing horizon or a regular origin. Then the metric of the space-time is given by}
\begin{equation}
ds^2 = - \left[1 - \frac{2 m(v,r)}{(N-3)r^{N-3}}\right] d v^2 + 2 \epsilon d v d r + r^2 (d \Omega_{N-2})^2,
\hspace{.3in} (\epsilon = \pm 1)\label{me}
\end{equation}
where
\begin{equation}
m(v,r) = \left\{ \begin{array}{ll}
        M(v)                      &   \hspace{.5in}    \mbox{if $C(v) = 0$}, \\
         & \\
       M(v) - 8 \pi C(v) \left(\frac{N-3}{N-2}\right) \frac{1}{(N-2)k + 1}r^{(N-2)k + 1}   &  \hspace{.5in}     \mbox{if $C(v)
\ne 0$ and $k \neq -1/(N-2)$}, \\
        & \\
       M(v) - 8 \pi C(v) \left(\frac{N-3}{N-2}\right) \ln r & \hspace{.5in}      \mbox{if $C(v) \neq 0$ and $k =
-1/(N-2)$}.
                \end{array}
        \right.
\label{eq:mv}
\end{equation}
\begin{equation} T^a_b =
\frac{C(v)}{r^{(N-2)(1-k)}} \mbox{{{diag}}}[1, 1, k,\ldots, k]. \label{emt}
\end{equation}
and
\begin{equation}
T^r_v = \left\{ \begin{array}{ll}
       \frac{1}{8 \pi r^{N-2}} \left(\frac{N-2}{N-3}\right) \frac{\partial M}{\partial v} -
       \frac{1}{(N-2)k + 1} \frac{\partial C}{\partial v } r^{(N-2)(k-1) + 1}   &  \hspace{.5in}
        \mbox{if $k \neq -1/(N-2)$}, \\
        & \\
       \frac{1}{8 \pi r^{N-2}} \left(\frac{N-2}{N-3}\right) \frac{\partial M}{\partial v} -
       \frac{1}{r^{N-2}}\frac{\partial C}{\partial v }\ln r
       & \hspace{.5in}      \mbox{if $k = -1/(N-2)$}.
                \end{array}
        \right.\label{emt2}
\end{equation}
\textit{Here, M(v) and C(v) are the arbitrary functions whose
values depend on the boundary conditions and the fundamental
constants of the underlying matter.} \\

\noindent {\bf Proof}: Expressed in terms of Eddington coordinate,
 the metric of general spherically symmetric space-time in
 $N$-dimensional space-times
 \cite{bi,pd,jr,ns} is,
\begin{equation}
ds^2 = - A(v,r)^2 f(v,r)\;  dv^2
 +  2 \epsilon A(v,r)\; dv\; dr + r^2 (d \Omega_{N-2})^2.
\label{eq:me2}
\end{equation}
Here $A(v,r)$ is an arbitrary function. It is useful to introduce
a local mass function $m(v,r)$ defined by $f(v,r) = 1 - {2
m(v,r)}/{(N-3)r^{(N-3)}}$. For $m(v,r) = m(v)$ and $A=1$, the
metric reduces to the $N$-dimensional Vaidya metric \cite{ns}.

In the static limit, this metric can be obtained from the metric
in the usual, spherically symmetric form,
\begin{equation}
ds^2 = -f(r)\; dt^2 + \frac{dr^2}{f(r)} + r^2 (d \Omega_{N-2})^2.
\end{equation}
by the coordinate transformation
\begin{equation}
dv = A(r)^{-1} ( dt + \epsilon \frac{dr}{f(r)} )
\end{equation}
In case of spherical symmetry, even when $f(r)$ is replaced by
$f(t,r)$, one can cast the metric in the form (\ref{eq:me2})
\cite{visser}.

The non-vanishing components of the Einstein tensor \cite{pd} are
\begin{subequations}
\label{fe1}
\begin{eqnarray}
&& G^v_r   = \frac{(N-2)}{r A^2} \frac{\partial A}{\partial r}, \label{equationa}
\\
&& G^r_v   = - \frac{(N-2)}{2 r} \frac{\partial f}{\partial v}, \label{equationb}
\\
&& G^v_v =  \frac{(N-2)}{2 r^2} \left[ r \frac{\partial f}{\partial r} - (N-3)(1-f) \right],
\label{equationc} \\
&& G^r_r =  \frac{(N-2)}{2 r^2} \left[ r \frac{\partial f}{\partial r} - (N-3)(1-f) \right] + \frac{(N-2)}{r
A}f \frac{\partial A}{\partial r},
\label{equationd} \\
&& 2 r^2 G^{\theta_i}_{\theta_i} = r^2 \frac{\partial^2 f}{\partial r^2} + (N-3)\left( 2 r \frac{\partial
f}{\partial r} - (N-4)(1-f)\right) +  2(N-3)\frac{r f}{A} \frac{\partial A}{\partial r} + 2
\frac{r^2}{A^2} \frac{\partial^2 A}{\partial v \partial r} \\
\nonumber && \hspace{0.56in}~+~ 3 \frac{r^2}{A} \frac{\partial f}{\partial r}\frac{\partial A}{\partial r} + 2
\frac{r^2 f}{A} \frac{\partial^2 A}{\partial r^2} - 2 \frac{r^2}{A^3} \frac{\partial A}{\partial r}
\frac{\partial A}{\partial v} , \label{equatione}
\end{eqnarray}
\end{subequations}
where $\{ x^a\} = \{v,\;r,\; \theta_1, \ldots \; \theta_{N-2} \}$. We shall consider the special case $T^v_r=0$
(hypothesis), which means from Eq.~(\ref{equationa}), $A(v,r)=g(v)$.  This also implies that $G^v_v=G^r_r$
(Eq.~(\ref{equationd})). However, by introducing another null coordinate $\overline{v} = \int g(v) dv$, we can
always set without the loss of generality, $A(v,r) = 1$. Hence, the metric takes the form,
\begin{equation}
ds^2 = - \left[1 - \frac{2 m(v,r)}{(N-3)r^{(N-3)}}\right] d v^2 + 2\epsilon d v d r + r^2 (d \Omega_{N-2})^2.
\label{metric}
\end{equation}
Therefore the entire family of solutions we are searching for is determined by a single function $m(v,r)$.
Henceforth, we adopt here a method similar to Salgado \cite{ms} which we modify here to accommodate the non
static case. In what follows, we shall consider $\epsilon =1$. The Einstein field equations are
\begin{equation}
R_{ab} - \frac{1}{2} R g_{ab} = 8 \pi T_{ab} ,\label{efe1}
\end{equation}
and combining Eqs.~(\ref{fe1}) and (\ref{efe1}), we have if $a
\neq b$, $T^a_b=0$ except for a non-zero off-diagonal components
$T^r_v$. In addition, we observe that $T_v^v = T_r^r$. Thus the
EMT can be written as :
\[
T^a_b = \left(%
\begin{array}{cccccc}
T^v_v \; & T^v_r\;  & 0 \;  & 0 \;  & . \;  & . \; \\

T^r_v & T^r_r & 0 & 0  & . \;  & . \; \\
   0 & 0 &T^{\theta_1}_{\theta_1}& 0  & . \;  & . \; \\
   . & . & . & .    & . \;& . \; \\
   . & . & . & .    & . \;&  T^{\theta_{N-2}}_{\theta_{N-2}}\; \\
\end{array}%
\right).
\]
which in general belongs to a Type II fluid with
$T^{\theta_1}_{\theta_1}=T^{\theta_2}_{\theta_2} =\, .\, .\, .\, =
T^{\theta_{N-2}}_{\theta_{N-2}}$ It may be recalled that EMT of a
Type II fluid has a double null eigen vector, whereas an EMT of a
Type I fluid has only one time-like eigen vector \cite{he}.
 On the other hand, from the Einstein equations,
it follows that
\begin{equation}
\nabla_a T^a_b=0. \label{de}
\end{equation}
Enforcing the conservation laws $\nabla_a T^a_b=0$, yields the following non-trivial differential equations:
\begin{equation}
\frac{\partial T^r_r}{\partial r} =  - \frac{(N-2)}{r}(T^r_r -
T^{\theta_1}_{\theta_1}), \label{divergence}
\end{equation}
and, using, $T^r_r = T^v_v$,
\begin{equation}
\frac{\partial T^v_v}{\partial v} = - \frac{\partial T^r_v}{\partial r} - \frac{(N-2)}{r}T^r_v.
\label{divergencea}
\end{equation}
Using the assumption made above that ${ T^{\theta_1}_{\theta_1} = kT^{r}_{r} }$, we obtain the following linear
differential equation
\begin{equation}
    \frac{\partial T^r_r }{\partial r} = - \frac{(N-2)}{r}(1-k) T^r_r,
      \label{diff}
\end{equation}
which can be easily integrated to give
\begin{equation}
T^r_r = \frac{C(v)}{r^{(N-2)(1-k)}}, \label{emts}
\end{equation}
where $C(v)$ is an arbitrary function of $v$, arising as an integration constant. Then, using hypothesis (iii), we conclude that
\begin{equation} T^a_b =
\frac{C(v)}{r^{(N-2)(1-k)}} \mbox{{{diag}}}[1, 1, k,\ldots, k]. \label{emtll}
\end{equation}
 Now using Eqs.~(\ref{fe1}) $\left[ \mbox{with~} f(v,r) = 1 - {2 m(v,r)}/{(N-3)r^{(N-3)}} \right]$,
(\ref{efe1}) and (\ref{emts}), we get
\begin{equation}
\frac{\partial m}{\partial r} = -8 \pi
\left(\frac{N-3}{N-2}\right) \frac{C(v)}{r^{-(N-2)k}},
\end{equation}
which trivially integrates to
\begin{equation}
m(v,r) = \left\{ \begin{array}{ll}
        M(v)                      &   \hspace{.5in}    \mbox{if $C(v) = 0$}, \\
         & \\
       M(v) - 8 \pi C(v) \left(\frac{N-3}{N-2}\right) \frac{1}{(N-2)k + 1}r^{(N-2)k + 1}   &  \hspace{.5in}     \mbox{if $C(v)
\ne 0$ and $k \neq -1/(N-2)$}, \\
        & \\
       M(v) - 8 \pi C(v) \left(\frac{N-3}{N-2}\right) \ln r & \hspace{.5in}      \mbox{if $C(v) \neq 0$ and $k =
-1/(N-2)$}.
                \end{array}
        \right.
\label{mvr}
\end{equation}
Here the function $M(v)$ arises as a result of integration. What
remains to be calculated is the only non-zero off-diagonal
component $T^r_v$ of the EMT. From Eqs.~(\ref{fe1}) and
(\ref{efe1}), one gets
\begin{equation}
T^r_v = \frac{1}{8 \pi r^{N-2}} \left(\frac{N-2}{N-3}\right) \frac{\partial m}{\partial v},
\end{equation}
which, on using Eq.~(\ref{mvr}), gives
\begin{equation}
T^r_v = \left\{ \begin{array}{ll}
       \frac{1}{8 \pi r^{N-2}} \left(\frac{N-2}{N-3}\right) \frac{\partial M}{\partial v} -
       \frac{1}{(N-2)k + 1} \frac{\partial C}{\partial v } r^{(N-2)(k-1) + 1}   &  \hspace{.5in}
        \mbox{if $k \neq -1/(N-2)$}, \\
        & \\
       \frac{1}{8 \pi r^{N-2}} \left(\frac{N-2}{N-3}\right) \frac{\partial M}{\partial v} -
       \frac{1}{r^{N-2}}\frac{\partial C}{\partial v }\ln r
       & \hspace{.5in}      \mbox{if $k = -1/(N-2)$}.
                \end{array}
        \right.
\end{equation}
It is seen that Eq.~(\ref{divergencea}) is identically satisfied.
Hence the theorem is proved. The theorem proved above represents a
general class of non-static, $N$-dimensional spherically symmetric
solutions to Einstein's equations describing radiating black-holes
with the EMT which satisfies the conditions in accordance with
hypothesis (iii).  The solutions generated here highly rely on the
assumption $(iii)$. On the other hand, although hypothesis (iv) is
not used a priori for proving the result, but it is indeed
suggested by regularity of the solution at the origin, from which,
$ T^v_v= T^r_r|_{r=0}$ (see \cite{ms} for further details).

The family of the $N$-dimensional solutions outlined here contains
$N$-dimensional version of, for instance, Vaidya \cite{ns,iv}
Bonnor-Vaidya \cite{pd,cbb}, dS/AdS \cite{pd}, global monopole
\cite{gm,pd,jr}, Husain \cite{vh,pd,jr}, and Harko-Cheng SQM
solution \cite{hc,ns}. Obviously, by proper choice of the
functions $M(v)$ and $C(v)$, and $k-$index, one can generate as
many solutions as required. The above solutions include most of
the known Vadya-based spherically symmetric solutions of the
Einstein field equations.  When $N=4$, the $4D$ solutions derived
in \cite{dg,ww} can be recovered. The static black holes
solutions, in both HD \cite{rcm} and in 4D \cite{ms}, can be
recovered by setting $M(v) = M, \; C(v) = C$ , with M and C as
constants, in which case matter is Type I.

 In summary, we have shown that the metric
 \begin{equation}
 ds^2 = -\left[1 - \frac{2M(v)}{(N-3)r^{N-3}} + \frac{16 \pi C(v)}{(N-2)[(N-2)k + 1]}
 r^{(N-2)(k-1)+2}\right] dv^2 + 2  dv dr + r^2
 (d\Omega_{N-2})^2,
 \end{equation}
is a solution of the Einstein equations for the  stress energy
tensor Eqs.~(\ref{emt}) and (\ref{emt2}).
A metric is considered to be asymptotically flat if in the vicinity
of a spacelike hypersurface its components behave as
\begin{equation}
g_{ab} \rightarrow \eta_{ab} + \frac {\alpha_{ab}(x^{c}/r,t)}{r} +
\mathcal{O}\left(\frac{1}{r^{1+\epsilon}}\right),
\end{equation}
as $r \rightarrow \infty$. ($\epsilon >0$, $\eta_{ab}$ is the
Minkowski metric, $\alpha_{ab}$ is an arbitrary symmetric tensor,
and $x^{c}$ is a flat coordinate system at spacelike infinity).
According to this definition, our metrics Eq.~(\ref{me}) are
asymptotically flat for $k
> -1/(N-2)$ and are cosmological for $k < -1/(N-2)$.  In particular, for k=-1, $M(v)=M$ and $ 2 C(v)=Q^2$,
the metric is just higher dimensional Reissner-Nordstr$\ddot{o}$m.
The detailed of the asymptotic structure of spatial infinity in higher-dimensional space-times can be found in Ref. \cite{st} and the different conformal diagrams for maximal extension of 4D Vaidya is discussed by Fayos {\it et al.} \cite{fps}.

The theorem proved shows that rather than a mathematical
coincidence, the above form of the metric is a consequence of the
features of the energy-momentum tensor considered. In the above
exact solutions, the associated energy momentum tensors share some
properties that are taken into account in the theorem in a general
fashion without specifying the nature of the matter. Therefore, the
theorem helps to characterize a whole two-parameter family of
solutions to the Einstein field equations.

The solutions discussed in the section are characterized by two
arbitrary functions $M(v)$ and $C(v)$, and the cosmological
constant  $\Lambda$. Thus one would like to generalize the above
theorem to include $\Lambda$. We can show that the energy momentum
tensor components, in general, can be written as, $T^a_b =
T^a_{(f)b} - \frac{\Lambda}{8 \pi}\delta^a_b$ \cite{ms,eg}, where
$\Lambda$ is the cosmological constant and $T^a_{(f)b}$ is energy
momentum tensor of the matter fields that satisfy
$T^{\theta}_{(f)\theta} = kT^r_{(f)r}$. A trivial extension of the
theorem allows one to cover a three-parameter family of solutions,
with one of the parameters
 being a cosmological constant $\Lambda$.  Next, we just state
 (proof being similar) the generalization of the Theorem  I. \\

 {\it{\bf Theorem - II}: Let ($M,~g_{ab}$) be an
N-dimensional space-time [sign$(g_{ab})$ = $+(N-2)$] such that (i)
It is non-static and spherically symmetric, (ii) it satisfies
Einstein field equations, (iii) the total energy-momentum tensor
is given by $T^a_b = T^a_{(f)b} - \frac{\Lambda}{8
\pi}\delta^a_b$, where $\Lambda$ is the cosmological constant and
$T^a_{(f)b}$ is energy momentum tensor of the matter fields, (iv)
in the Eddington coordinates where $ds^2 = - A(v,r)^2 f(v,r)\;
dv^2
 +  2 A(v,r)\; dv\; dr + r^2 (d \Omega_{N-2})^2$, the
EMT $T^a_{(f)b}$ satisfies the conditions $T^v_{(f)r}=0$, $T^{\theta_1}_{(f)\theta_1} = kT^r_{(f)r}$, $(k =
\mbox{const.}\; \in \; \mathcal{R}$), (v) it possesses a regular Killing horizon or a regular origin. Then the
metric of the space-time is given by metric (\ref{me})}, where
\begin{equation}
m(v,r) = \left\{ \begin{array}{ll}
        M(v) + \frac{(N-3)}{(N-2)(N-1)} \Lambda r^{N-1}                      &   \hspace{.5in}
\mbox{if $C(v) = 0$}, \\
         & \\
       M(v) - 8 \pi C(v) \left(\frac{N-3}{N-2}\right) \frac{1}{(N-2)k + 1}r^{(N-2)k + 1} + \frac{(N-3)}{(N-2)(N-1)} \Lambda r^{N-1}   &
\hspace{.5in}     \mbox{if $C(v) \ne 0$ and $k \neq -1/(N-2)$}, \\
         & \\
       M(v) - 8 \pi C(v) \left(\frac{N-3}{N-2}\right) \ln r + \frac{(N-3)}{(N-2)(N-1)} \Lambda r^{N-1}& \hspace{.5in}      \mbox{if
$C(v) \neq 0$ and $k = -1/(N-2)$}.
                \end{array}
        \right.                         \label{eq:mvl}
\end{equation}
\begin{equation}
T^a_b =  \frac{C(v)}{r^{(N-2)(1-k)}} {\mbox{diag}}[1, 1, k,\ldots,
k] - \frac{\Lambda}{8 \pi} {\mbox{diag}}[1, 1, 1,\ldots, 1]
\label{emtl}
\end{equation}
and
\begin{equation}
T^r_v = \left\{ \begin{array}{ll}
       \frac{1}{8 \pi r^{N-2}} \left(\frac{N-2}{N-3}\right) \frac{\partial M}{\partial v} -
       \frac{1}{(N-2)k + 1} \frac{\partial C}{\partial v } r^{(N-2)(k-1) + 1}   &  \hspace{.5in}
        \mbox{if $k \neq -1/(N-2)$}, \\
        & \\
       \frac{1}{8 \pi r^{N-2}} \left(\frac{N-2}{N-3}\right) \frac{\partial M}{\partial v} -
       \frac{1}{r^{N-2}}\frac{\partial C}{\partial v }\ln r
       & \hspace{.5in}      \mbox{if $k = -1/(N-2)$}.
                \end{array}
        \right.\label{emtl1}
\end{equation}
\textit{Here, M(v) and C(v) are arbitrary functions of $v$, arising as integration constants, whose values depend on the boundary conditions and the fundamental constants of
the underlying matter.} \\

\section{Energy Conditions}
The family of solutions discussed here, in general, belongs to Type II fluid defined in \cite{he}. When
$m=m(r)$, we have $\mu$=0, and the matter field degenerates to type I fluid \cite{ww}. In the rest frame
associated with the observer, the energy-density of the matter will be given by (assuming $\Lambda = 0$),
\begin{equation}
\mu = T^r_v,\hspace{.1in}\rho = - T^t_t = - T^r_r = -\frac{C(v)}{r^{(N-2)(1-k)}}, \label{energy}
\end{equation}
 and the principal pressures are $P_i =
T^i_i$ (no sum convention).  Therefore $P_r = T^r_r = - \rho$ and
$P_{\theta_1} = k P_r = -k \rho$ (hypothesis $(iii)$). \\

 \noindent \emph{a) The weak energy
conditions} (WEC): The energy momentum tensor obeys inequality $T_{ab}w^a w^b \geq 0$ for any timelike vector,
i.e.,
\begin{equation}
\mu \geq 0,\hspace{0.1 in}\rho \geq 0,\hspace{0.1 in} P_{\theta_1}
= P_{\theta_2}= .\, .\, .\, = P_{\theta_{(N-2)}} \geq 0.
\label{wec}
\end{equation}
We say that strong energy condition (SEC), holds for Type II fluid if, Eq.~(\ref{wec}) is true., i.e., both WEC
and SEC, for a Type
II fluid, are identical. \\

\noindent {\emph{b) The dominant energy conditions }}: For any timelike vector $w_a$, $T^{ab}w_a w_b \geq 0$,
and $T^{ab}w_a$ is non-spacelike vector, i.e.,
\begin{equation}
\mu \geq 0,\hspace{0.1 in}\rho \geq P_{\theta_1}, P_{\theta_2}=
.\, .\, .\, = P_{\theta_{(N-2)}} \geq 0. \hspace{0.1 in}
\end{equation}
Clearly, $(a)$ is satisfied if $C(v)\leq 0, k\leq 0$. However, $\mu
> 0$ gives the restriction on the choice of the functions $M(v)$
and $C(v)$. From Eq.~(\ref{emt2}), $( k \neq -1/(n-2) ),$ we
observe $\mu
> 0$ requires,
\begin{equation}
\frac{1}{8 \pi r^{N-2}} \left(\frac{N-2}{N-3}\right) \frac{\partial M}{\partial v} -
       \frac{1}{(N-2)k + 1} \frac{\partial C}{\partial v } r^{(N-2)(k-1) + 1} > 0.
\end{equation}
This, in general, is satisfied, if
\begin{equation}\label{ecM}
\left(\frac{N-2}{N-3}\right) \frac{\partial M}{\partial v}
> 0, \mbox{ and, either }\frac{\partial C}{\partial v} > 0
\mbox{ and } k < -1/(N-2), \mbox{ or }\frac{\partial C}{\partial v} < 0 \mbox{ and } k > -1/(N-2).
\end{equation}
\noindent On the other hand, for $ k = -1/(N-2) $, $\mu \geq 0$ if ${\partial M}/{\partial C} \geq 8 \pi
\left(\frac{N-3}{N-2}\right) \ln r$. The DEC holds if $C(v) \leq 0$  and  $-1 \leq k \leq 0$, and the function
$M$ is subject to the condition (\ref{ecM}).  Clearly,  $0 \leq -k \leq 1$.

\section{Singularity and Horizons}
The invariants are regular everywhere except at the origin $r = 0$, where they diverge.  Hence, the space-time
has the scalar polynomial singularity \cite{he} at $r=0$.  The nature (a naked singularity or a black hole) of
the singularity can be characterized by the existence of radial null geodesics emerging from the singularity.
The singularity is at least locally naked if there exist such geodesics, and if no such geodesics exist, it is
a black hole. The study of causal structure of the space-time is beyond the scope of this paper and will be
discussed elsewhere.

In order to further discuss the physical nature of our solutions, we introduce their kinematical parameters.
Following York \cite{jy} a null-vector decomposition of the metric (\ref{me}) is made of the form
\begin{equation}\label{gab}
g_{ab} = - n_a l_b - l_a n_b + \gamma_{ab},
\end{equation}
where,
\begin{subequations}
\label{nv}
\begin{eqnarray}
n_{a} = \delta_a^v, \: l_{a} = \frac{1}{2} \left[ 1  - \frac{2 m(v,r)}{(N-3)r^{N-3}} \right ] \delta_{a}^v +
\delta_a^r, \label{nva}
 \\
\gamma_{ab} = r^2 \delta_a^{\theta_1} \delta_b^{\theta_1} + r^2
\left[\left( \prod_{j=1}^{i-1} sin^2({\theta}_j) \right) \right]
\delta_a^{\theta_i} \delta_b^{\theta_i}, \label{nvb}
\\
l_{a}l^{a} = n_{a}n^{a} = 0 \; ~l_a n^a = -1, \nonumber \\ l^a \;\gamma_{ab} = 0; \gamma_{ab} \; n^{b} = 0,
\label{nvd}
\end{eqnarray}
\end{subequations} with $m(v,r)$ given by Eq.~(\ref{eq:mv}).  The
optical behavior of null geodesics congruences is governed by the
Carter \cite{bc} form of the Raychaudhuri equation
\begin{equation}\label{re}
   \frac{d \Theta}{d v} = \mathcal{K} \Theta - R_{ab}l^al^b-
   (\gamma^c_c)^{-1}
   \Theta^2 - \sigma_{ab} \sigma^{ab} + \omega_{ab}\omega^{ab},
\end{equation}
with expansion $\Theta$, twist $\omega$, shear $\sigma$, and
surface gravity $\mathcal{K}$. Here $R_{ab}$ is the
$N$-dimensional Ricci tensor, $\gamma_c^c$ is the trace of the
projection tensor for null geodesics. The expansion of the null
rays parameterized by $v$ is given by
\begin{equation}\label{theta}
\Theta = \nabla_a l^a - \mathcal{K},
\end{equation}
where the $\nabla$ is the covariant derivative. In the present case, $\sigma = \omega = 0$ \cite{jy}, and the
surface gravity is,
\begin{equation}\label{sg}
\mathcal{K} = - n^a l^b \nabla_b l_a.
\end{equation}
Spherically symmetric irrotational space-times, such as under
consideration, are vorticity and shear free. The structure and
dynamics of the horizons are then only dependent on the expansion,
$\Theta$. As demonstrated by York \cite{jy}, horizons can be
obtained by noting that (i) apparent horizons are defined as
surface such that $\Theta \simeq 0$ and (ii) event horizons are
surfaces such that $d \Theta /dv \simeq 0$. Substituting
Eqs.~(\ref{eq:mvl}), (\ref{nv}) and (\ref{sg}) into
Eq.~(\ref{theta}), we get, ($k \ne -1/2$)
\begin{equation}\label{ah1}
\Theta= \frac{1}{r} \left[1 - \frac{2 \overline{M}(v)}{r^{N-3}} + Q^2(v) r^{(N-2)k-(N-4)} - \chi^2 r^2 \right],
\end{equation}
where,
\begin{equation}\label{ah11}
\overline{M}(v)=\frac{M(v)}{(N-3)}\;\;\;,\chi^2 = \frac{2 \Lambda}{(N-2)(N-1)}\;\;\;,\;\;\; Q^2(v)=\frac{16 \pi
C(v)}{(N-2)((N-2)k+1)}.
\end{equation}Since the York conditions require that at apparent horizons $ \Theta$ vanish, it follows form the
Eq.~(\ref{ah1}) that apparent horizons will satisfy
\begin{equation}\label{ah2}
\chi^2r^{N-1} - Q^2 r^{(N-2)k+1} - r^{N-3} + 2\overline{M}(v),=0,
\end{equation}
which in general has two positive solutions. For $\chi^2 =Q^2= 0$, we have Schwarzschild horizon
$r=(2\overline{M})^{\frac{1}{N-3}}$, and for $M=Q^2=0$ we have de Sitter horizon $r=1/ \chi$.  As mentioned
above, for $k=-1$, one gets Bonnor-Vaidya solution, in which case the various horizons are identified and
analyzed by Mallett \cite{rm} and hence, to conserve space, we shall avoid the repetition of same. For general
$k$, as it stands, Eq.~(\ref{ah2}) will not admit simple closed form solutions.  However, for
\begin{equation}\label{ahq}
Q^2 = Q^2_c = \frac{-(N-3)}{( (N-2)k +1)} \left[
\frac{2M((N-2)k+1)}{(N-3)((N-2)k-(N-4))}\right]^{\frac{(N-4)-(N-2)k}{(N-3)}},
\end{equation}
with $\chi^2 = 0$, the two roots of the Eq.~(\ref{ah2}) coincide and there is only one horizon
\begin{equation}\label{ahs}
  r = \left[
\frac{2M((N-2)k+1)}{(N-3)((N-2)k-(N-4))}\right]^{\frac{1}{(N-3)}}
\end{equation}
  For $Q^2
\leq Q^2_c$ there are two horizons, namely a cosmological horizon and a black hole horizon.  On the other hand
if, the inequality is reversed, $Q^2 > Q^2_c$ no horizon would form.

\section{Concluding remarks}
In the study of the Einstein equations in the 4D space-time
several powerful mathematical tools were developed, based on the
space-time symmetry, algebraical structure of space-time, internal
symmetry and solution generation technique, global analysis, and
so on. It would be interesting how to develop some of these
methods to higher dimensional space-time. With this as motivation,
plus the fact that exact solutions are
 always desirable and valuable, we have extended to higher dimensional space-time, a recent
theorem \cite{dg} and it's trivial extension (that includes
cosmological term $\Lambda$), which, with certain restrictions on
the EMT, characterizes a large family of radiating black hole
solutions in N-dimensions, representing, in general, spherically
symmetric Type II fluid.  In particular, the monopole-de
Sitter-charged Vaidya and Husain solutions can be generated form
our analysis and when $n=4$, one recovers the 4D black solutions.
If $M = C = $ constant, we have $\mu$=0, and the
matter field degenerates to type I fluid and we can generate
static black hole solutions by a proper choice of these constants.
Since many known solutions are identified as particular case of
this family and hence it would be interesting to ask whether there
exist realistic matter that follows the restrictions of the
theorem that would generate a new black hole solution.

The solutions depend on one parameter $k$, and two arbitrary
functions $M(v)$ and $C(v)$ (modulo energy conditions). It is
possible to generate various solutions by proper choice of these
functions and parameter $k$. Further, most of the known static
spherically symmetric black hole solutions, in 4D and HD, can be
recovered from our analysis.

It should be interesting to apply these metrics to study the gravitation collapse and naked singularities formation.
Finally, the result obtained would also be relevant in the context
of string theory which is often said to be next "theory of
everything" and in the study of gravitational collapse.

\acknowledgements Authors are
grateful to the referee for constructive criticism. One of the author(SGG) would like to thank IUCAA, Pune for
hospitality while this work was done.

\noindent

\end{document}